\documentclass[aps,pre,twocolumn,superscriptaddress,groupedaddress]{revtex4}  

\usepackage{graphicx} 
\usepackage{dcolumn}   
\usepackage{bm}        
\usepackage{amssymb}   
\usepackage{amsmath}
\usepackage{color,soul}
\usepackage{xcolor}
\usepackage{makecell}

\newcommand{\norm}[1]{\left\lVert#1\right\rVert}

\begin{document} 
\widetext
\title{Optimal Partitions for Nonparametric Multivariate Entropy Estimation}

\author{Z.~Keskin} 
\affiliation{Financial Computing \& Analytics, University College London, Gower Street, WC1E 6EA, London, United Kingdom}
\vskip 0.25cm
\date{\today}

\begin{abstract}

  Efficient and accurate estimation of multivariate empirical probability distributions is fundamental to the calculation of information-theoretic measures such as mutual information and transfer entropy. 
  Common techniques include variations on histogram estimation which, whilst computationally efficient, are often unable to precisely capture the probability density of samples with high correlation, kurtosis or fine substructure, especially when sample sizes are small.
  Adaptive partitions, which adjust heuristically to the sample, can reduce the bias imparted from the geometry of the histogram itself, but these have commonly focused on the location, scale and granularity of the partition, the effects of which are limited for highly correlated distributions. 

  In this paper, I reformulate the differential entropy estimator for the special case of an equiprobable histogram, using a k-d tree to partition the sample space into bins of equal probability mass. By doing so, I expose an implicit rotational orientation parameter, which is conjectured to be suboptimally specified in the typical marginal alignment. I propose that the optimal orientation minimises the variance of the bin volumes, and demonstrate that improved entropy estimates can be obtained by rotationally aligning the partition to the sample distribution accordingly. Such optimal partitions are observed to be more accurate than existing techniques in estimating entropies of correlated bivariate Gaussian distributions with known theoretical values, across varying sample sizes (99\% CI).

\end{abstract}

\maketitle

\section{Introduction}

    Information theory presents a rich field for novel approaches in statistical methods, and a considerable literature now exists discussing the limitations of classical statistical models, such as Pearson correlation and Spearman's rank correlation, and contrasting these with models able to capture nonlinear relationships, such as Reshef et al.'s maximal information coefficient \cite{reshef2011MIC}, or to distinguish directionality, like Schreiber's transfer entropy \cite{schreiber2000measuring}. 

    The preponderance of these information-theoretic techniques rely on accurate estimates of the empirical probability distributions from sample data. In the absence of a theoretical prior distribution, the estimation is described as nonparametric. 
    The first, simplest and most computationally efficient approach to nonparametric probability density estimation is the histogram. In histogram estimation the sample space is partitioned into disjoint, usually hyperrectangular, subspaces commonly referred to as ``bins", which are aligned along each axis and typically of equal width. The estimated population density in each bin is calculated as the proportion of total samples observed, scaled by the proportional volumes of each bin.

    In effect, the number of bins is an input parameter which is selected to account for the bias-variance trade-off; too few bins will fail to capture the substructure in enough granularity, whilst too many bins will result in information-sparse histograms containing large numbers of empty bins. Investigations into suitable parameters for histogram partitions have developed various guidelines based in general on sample size. Sturges developed the first formal approach for setting this parameter, suggesting the appropriate number of bins scaled with the logarithm of the sample size ($B = \log_2{N}+1$). Scott provided a more rigorous recommendation, minimising the integrated mean square error of the density estimate, to derive the optimal data-based bin width for the histogram \cite{scott1979optimal}. Scott noted in the asymptotic case that the optimal width scales inversely with cubic $N$, and linearly with the density function, hence suggesting $h^{*}=3.49 \sigma N ^{-\frac{1}{3}}$, based on Gaussian assumptions, where $\sigma$ is the standard deviation of the univariate data. However, these equations were derived only for univariate histograms with equal-width bins \cite{sturges1926choice}. 
    
    Freedman and Diaconis developed this further, suggesting the use of the interquartile range, rather than the standard deviation, to be more robust to distributions of high skew and kurtosis \cite{freedman1981histogram}. Knuth adopted a Bayesian approach, defining the partition in terms of a piecewise-constant density estimator, and maximising the posterior probability to select an optimum number of bins \cite{knuth2006optimal}.  
    
    Each of these approaches were presented for equal-width partitions, although it was noted by both Knuth and Scott that the density estimates may be improved by allowing the bins to vary in size, since guidelines for parameter selection applied to equal-width bins cannot truly account for the local effects of the distribution. For example, probability distributions with fine substructure, bimodality or high degrees of kurtosis, will suffer more acutely from the bias-variance trade-off. This trade-off is exacerbated in higher dimensions, known as the curse of dimensionality, since sample size must increase exponentially to return the same sample density. This has practical consequences. For example, calculating the information-theoretic measure of conditional mutual information requires a minimum of three dimensions to estimate the multivariate joint entropy, which motivates improvements over naive partitioning.
    
    In addition to sample size, some partitioning routines incorporate features of the sample distribution, such as location and scale statistics; as well as selecting optimal parameters for the histogram partition, improved estimates can be obtained by adjusting the size of bins according to the observed sample density. Such partitions are described as adaptive, and can be achieved through a number of approaches. They are generally more computationally expensive, as the volume and position of each bin must be calculated in addition to counting the samples in each bin. 

    A relatively simple approach proceeds by partitioning the sample space into quantiles along each dimension independently. This technique is known as marginal equiquantisation, and has been used occasionally in the literature, for example by Palus for the estimation of entropy rates of complex time series \cite{paluvs1995testing}\cite{paluvs1996coarse}, and also by the author of this paper, in a recent investigation into the causal relationship between social media sentiment and cryptocurrency prices \cite{KeskinAste2020}. Though such approaches are computationally appealing, and can capture more information about the underlying distributions compared to naive equipartitions, they may fail to accurately capture the joint probability density since in general the product of marginal quantile partitions is not equal to the joint  (although the equivalence holds asymptotically in the limit of increasing granularity). Given the histogram density estimator assumes uniform density within each bin, the global estimator is non-smooth, except in the case where the global and local densities are equal. It is possible to ensure this by appropriate partitioning of the sample space, and such partitions are described as equiprobable.

    Equiprobable partitions have been used in the literature for the estimation of entropy, and there exist diverse algorithms to produce them. Hlav{\'a}{\v{c}}kov{\'a}-Schindler et al. is recommended for an introductory overview of the space \cite{hlavavckova2007causality}. Early success was pioneered by Fraser and Swinney \cite{fraser1986independent}, who developed an algorithm for partitioning a 2D sample space into equiprobable quadrants, which is applied recursively until local $\chi^2$ tests no longer indicate substructure. This approach anticipated progress in information-theoretic perspectives on causality, having been applied to calculate mutual information between a time series vector and a lagged vector of the same time series. However, without extending the approach to higher dimensions, the Fraser-Swinney algorithm is limited in its applicability; for example the calculation of transfer entropy using this approach would require the partitioning of a three-dimensional sample space.

    Cellucci et al. \cite{cellucci2005statistical} presented an adaptation to the Fraser-Swinney algorithm which applies a $\chi^2$ criterion to the overall probability distribution, rather than only to local subdivisions, resulting in the advantage of providing a global test of statistical independence. The Cellucci approach also results in fewer subdivisions, improving calculation time by two orders of magnitude, while estimating a numerical error within 4\%. 

    Darbellay and Vajda \cite{darbellay1999estimation} presented the definitive implementation of the equiprobable partition for mutual information, proving that the algorithm can estimate the mutual information in expectation asymptotically to arbitrary precision, even in higher dimensions. They also demonstrated the performance, in terms of bias and variance, against marginally-equiquantised and uniform-width partitions benchmarked against a parametric estimator of the known (bivariate Gaussian) distribution.

    Whilst previous approaches to equiprobable binning have delivered sample space partitions with uniform probability densities, these have not optimised in general the bias-variance trade off for a given coarseness in the partition. This is largely due to such partitions proceeding with a preferred orientation, and dividing the sample space along the marginal dimensions. However, in the case of entropy estimation, it can be shown that only the volume of each bin is required; it is permitted to orient the sample space using any rotation which is sub-space invariant. There is no \textit{a priori} reason for the sample space to be partitioned along the marginals and this paper conjectures that the optimal orientation is not in general along the marginals. 
    
    This paper proposes a novel improvement to the estimation of entropy using equiprobable partitions, by adapting the rotational alignment of partition marginals to the sample distribution, hence allowing coarser-grained partitions over the same sample space. This can improve the information density without sacrificing against bias or variance - in effect a partial remedy to the curse of dimensionality. To the author's knowledge, the approach of rotational alignment of partitions for improved entropy estimation is new, and the algorithm presents a valuable addition to the practitioner's repertoire for calculating any entropy-based measure. 

    The rest of the paper is organised as follows. Section \ref{s.background} provides the theory behind information-theoretic measures, and demonstrates the applicability of equiprobable partitions to estimating these by deriving the Shannon entropy exclusively in terms of relative bin volumes. The mathematical apparatus required for rotations in multivariate sample spaces is also discussed. Section \ref{s.method} presents the algorithm, and discusses key considerations around practical usage and appropriate parameterisation. The method to optimally align the partition is also demonstrated. Following this, in Section \ref{s.validation} we compare the performance of the optimal partition against naive and marginal equiquantised partitions, by generating randomised samples from bivariate Gaussian distributions and comparing the entropy estimates to the expected theoretical values.

\section{Background} \label{s.background}  

  The Shannon entropy for a multivariate continuous random variable is given by:

  \begin{eqnarray}
    \label{eq:entropy}
    H(\textbf{X}) = - \int_{-\infty}^{+\infty}  { p(\textbf{x}) \log{p(\textbf{x})}  } d\textbf{x},
  \end{eqnarray}

  where $\textbf{X} \in \mathbb{Re}^d $ is a multivariate random variable and $p(\textbf{x})$ is the probability density at position $\textbf{X} = \textbf{x}$. The entropy equation was first introduced by Gibbs, and subsequently discovered by Shannon as a measure of information \cite{shannon1948}. The information-theoretic identities in this section are stated with reference to Cover \& Thomas \cite{cover1999elements}. 

  First, define a sample space $\Omega \in \mathbb{Re}^d$, describing the domain of observed samples which is to be partitioned into $B$ bins, creating subspaces $\Omega_i$ such that:
  
  \begin{equation}
    \label{eq:sample-spaces}
    \Omega = \Omega_1 \cup \Omega_2 \cup ... \cup \Omega_B .
  \end{equation}

  and 
  \begin{equation}
    \label{eq:sample-spaces-intersection}   
    \Omega_i \cap \Omega_j = \O \quad \operatorname{if} i \neq j.
  \end{equation}

  Also define an operator, $v(\cdot)$, which returns the volume of $\Omega_i$ in the real sample space, such that $ \sum v(\Omega_i) = v(\Omega)$. A second operator  $n(\cdot)$ describes the number of samples in the subspace, such that $ \sum n(\Omega_i) = n(\Omega) = N$.

  Given sample data $\{\textbf{x}_0,\textbf{x}_1, ..., \textbf{x}_N \}$, the uncertainty of the distribution can be estimated from the partitioned sample space by calculating the entropy of the probability density estimator using a histogram approach. Such a discretisation restricts the permissible values of $\textbf{x}$ to a finite alphabet, dictated by the coarseness of the partition. One may write $\textbf{x}_i \in \{\textbf{x}_1, \textbf{x}_2, ... , \textbf{x}_B\}$, denoting the equivalence of the alphabet of  $\textbf{x}_i$ and the set of subspaces, $\Omega$.
  
  The entropy estimate of the partitioned sample space is given by:

  \begin{equation}
    \label{eq:entropy-discrete}
    H(\textbf{X}) = - \sum_{i=1}^{B} \hat{p}(\textbf{x}_i) v(\Omega_i) \log \hat{p}(\textbf{x}_i),
  \end{equation}

  where $\hat{p}(\textbf{x}_i)$ is the estimate of the probability density function $p(\textbf{x})$ of the bin $\Omega_i$ containing $\textbf{x}$, i.e. $p(\textbf{x} \in \Omega_i)$. This is estimated from:

  \begin{equation}
    \label{eq:probability-estimate}
    \hat{p}(\textbf{x}_i) = \frac{n(\Omega_i)}{N v(\Omega_i)},
  \end{equation}

  where $n(\Omega_i)$ is the number of samples in the bin containing sample $\textbf{x}_k$, $N$ is the total number of samples, and $B$ is the total number of bins in the partition. The estimate for entropy can therefore be written as:
  
  \begin{equation}
    \label{eq:entropy-estimate}
    H(\textbf{X}) \simeq - \sum_{i=1}^{B}  \frac{n(\Omega_i)}{N} 
      \log\left({ \frac{n(\Omega_i)}{N v(\Omega_i)} }\right).
  \end{equation}

  In simple histogram partitions, equal-sized bins mean the $v(\Omega_i)$ terms are constant. In equiprobable partitions, bins are arranged such that each contains as close to an equal proportion of samples as possible. In the ideal case where $\frac{n}{N} \in \mathbb{N}$, all bins share an equal probability mass of $m = \frac{n}{N}$, where $ n(\Omega_i) = n \; \forall \, k $ and so, by contrast, the bin volumes vary but the probability masses are constant. This implies:
  
  \begin{equation}  
    \label{eq:prob-mass} 
      \frac{n(\Omega_i)}{N} = \frac{1}{B} \qquad \forall i.
  \end{equation}
  
  Equiprobable partitions can be constructed using recursive binary trees. In these cases, $B$ is given by:

  \begin{equation}   
    \label{eq:numBins}
    B = 2^{(s \times d)}
  \end{equation}

  where $s$ is the depth of recursion, and $d$ is the dimensionality of the sample space.

  Under such a recursive binary partition, if follows from equation (\ref{eq:entropy-estimate}) that the estimator for entropy simplifies to: 

  \begin{equation}   
    \label{eq:equiprobable-entropy}
    H(\textbf{X}) \simeq \left( \frac{1}{2} \right)^{s \times d} \log_{2} \left[ \prod_{i=1}^{B} v(\Omega_i) \right] + s \times d
  \end{equation}

  from which it is clear that the information is able to be captured entirely from the relative volumes of the bins.
  
  This result is useful because any affine transformation does not impact the entropy of the sample, but can allow for an adaptive partition with reduced bias, by maximising the information for a given number of bins. The equiprobable partition also avoids the discontinuities in probability density estimates which are generally present in typical histogram estimates. The method further benefits from computational efficiency, as constructing the histogram encodes the entropy implicitly in the volumes of the bins, and this requires only the calculation of the k-d tree, which is bounded by $\mathbb{O}(N \log N)$ \cite{stowell2009fast}. 

In the following sections, we make use of equation (\ref{eq:equiprobable-entropy}) to calculate entropy estimates of joint distributions, by rotating the sample with respect to an equiprobable partition, and then compare these against classical histogram methods in estimating theoretical entropies.

Rotation operations are well defined in linear algebra in two dimensions. For the sample vector $\textbf{X}$, rotating the partition is equivalent to rotating the points around the barycentre. This can be achieved by first translating this centre point to the origin, and applying the matrix operation:

  \begin{equation}
    \label{eq:rotation2d}
    \textbf{X}^{\prime} = 
    \begin{pmatrix}
      \cos \theta & -\sin \theta \\
      \sin \theta & \cos \theta
  \end{pmatrix}
  \textbf{X},
  \end{equation}

where $\theta$ is the angle of rotation and $\textbf{X}^{\prime}$ is the rotated variable vector.

In three dimensions, rotation proceeds around an axis of rotation, which itself has an angular orientation with respect to the basis. Hence the rotation operation can be considered a sequence of axial rotations, and rotation matrices can be constructed sequentially. The non-commutativity of matrix multiplication requires the proper order of operations, which contributes to the complexity of rotation matrices for three-dimensional rotations. In practice, alternative methods are often used to represent rotations in dimensions of three and greater, although these may be considered less intuitive than rotation matrix techniques. Euler angles and unit quaternions represent sparse alternatives to describing rotations; unit quaternions require four parameters to specify a rotational alignment in three dimensions. Diebel \cite{diebel2006representing} and Terzakis et al. \cite{terzakis2018modified} are recommended as a reference on the various mathematical formulations of rotational orientation in three dimensions. The mathematics for higher-dimensional rotations beyond three dimensions is somewhat obscure in the literature but well understood; derivations are presented in Hanson \cite{hanson2011rotations}, and Hoffman et al. \cite{hoffman1972generalization}.

Modified Rodrigues Parameters (MRPs) represent the sparsest possible representation of the three-dimensional rotation operation, and are in the author's view the most intuitive technique for describing rotations generally in two or three dimensions. In the MRP formulation the rotation is described by a three-vector pointing along the axis of rotation, whose magnitude encodes the angle of rotation using the relationship:

\begin{equation}
    \label{eq:mrp}
  \norm{\textbf{R}} = \tan \left( \frac{\theta}{4} \right),
  \end{equation}

 where $\textbf{R} \in \mathbb{Re}^3$ is the vector defining the axis rotation, and $\theta \in [0,2\pi)$ is the angle in radians by which the partition is rotated. In two dimensions the rotation reduces to an MRP rotation around the $z$-axis. 

In addition to its parsimony, the MRP formulation is attractive for numerical optimisation, as it allows for interpolation and avoids the issue of gimbal lock present with Euler angles. Hence MRPs can be relatively simply exploited to calculate the optimal rotational alignment, and thereby align the orientation of the partition with the sample data; this technique was used for all rotations in this paper.

\section{Partition Method} \label{s.method}

To apply the formulae developed in Section \ref{s.background}, one must be able to algorithmically partition any sample of data $\textbf{X}$ in the sample space $\textbf{X} \in \Omega$ such that each bin contains the same number of samples. In this section is presented a computationally efficient nonparametric approach which generalises naturally to higher dimensions by adopting the concept of k-d trees, which are a subset of binary space partitioning algorithms \cite{bentley1975multidimensional}.
The procedure operates in a loop, executing binary partitions by dividing the sample space along the marginal median of each dimension. Recursion is employed to a pre-determined stopping depth $s \in {\mathbb{N}}$, where all dimensions are bisected once in each iteration of the recursion. Partitioning all dimensions in this way reduces bias that would otherwise be imparted by bisecting the dimensions unequally. The number of bins scales exponentially with dimensionality and depth (see equation (\ref{eq:numBins})), and so generally the appropriate depth parameter scales logarithmically with $N$.

\begin{figure}
  \centering
  \includegraphics[width=\columnwidth]{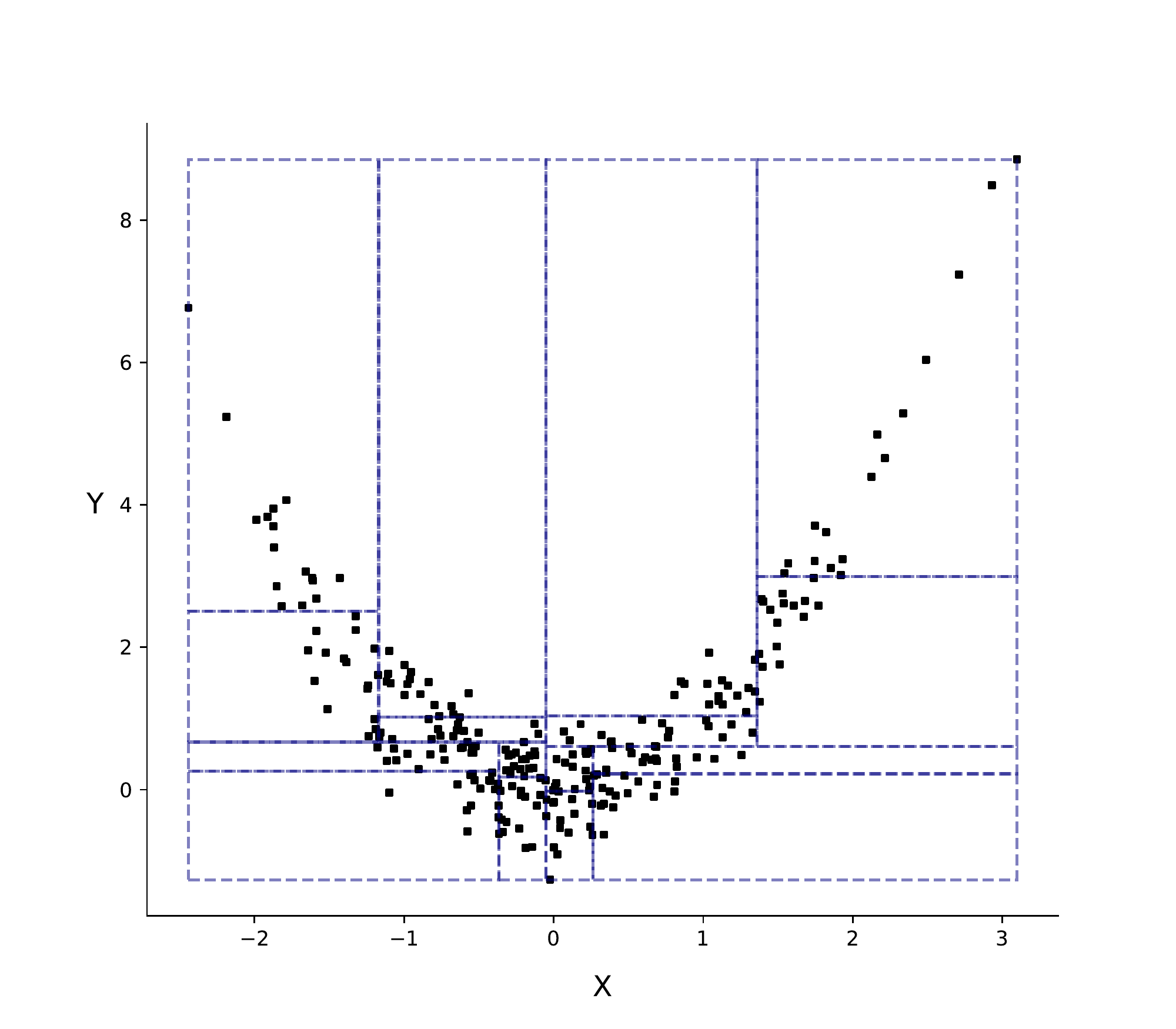}
  \caption{Representation of an equiprobable partition calculated using a recursive binary tree for a random sample of 256 data from $Y = X^2 + \mathcal{N} (0,0.5)$, where $X \sim \mathcal{N} (0,1)$. The regions of high probability density have bins of smaller volume and the regions of lower probability density have larger-volume bins; this is sufficient to estimate the entropy of the distribution.}
  \label{fig.partition}
\end{figure}

It is noted here the similarity of this partition algorithm to that of Stowell and Plumbley \cite{stowell2009fast}, who also propose a recursive binary partition using a k-d tree approach. In contrast to the algorithm used in this paper, which by necessity divides all subpartitions of sample space to an equal depth, Stowell and Plumbley proposed a data-driven stopping criterion, to determine the depth of branching to reach each leaf node independently; this does not result in general with an equiprobable partition, which is required for rotational alignment of the partition with the sample data.

The results of performing a recursive k-d tree partition without rotational alignment, to a stopping depth of $s=2$, over the sample space of a synthetic parabolic distribution is presented in figure \ref{fig.partition}. It is clear from the tails of the distribution that the second-order relationship between $X$ and $Y$ causes the partition volumes to be considerably larger in the top-left and top-right bins, where the probability density is lower. By contrast, near the stationary point, the samples are more densely clustered and bin volumes are smaller. Evidently, the variance in bin volumes encodes information from the distribution.

Careful thought is sufficient to determine that distributions of greater joint entropy correspond to smaller variances in bin volumes; in the limit of a uniform distribution, the expectation of the variance is zero. This may be considered analogous to the fact that, for a set of fractions summing to 1, the maximum product is when these fractions are equal. The analogy corresponds directly to equation (\ref{eq:equiprobable-entropy}), and it also motivates the choice of an objective function, which is used to define optimality of the partition's rotational orientation. It is desirable that the partition succeed in capturing information about the distribution whilst minimising the introduction of additional information. Defining the rotational orientation of the equiprobable partition as a parameter, this should be optimised to find the partition which imparts the least additional information to the entropy estimate. Specifically I define the partition as optimal which minimises the variance of bin volumes.

\subsection{Optimal Alignment}

To find the optimal partition alignment represents a constrained optimisation problem which, with reference to the MRP formulation, may be stated in the general case by:

\begin{equation}
  \label{eq:optimisation}
  \begin{aligned}
  & \underset{\textbf{R}}{\text{minimize}}
  & & \sum_{i=1}^B \mathbb{E} [v(\Omega^{\prime}_i)^2] - \mathbb{E}[v(\Omega^{\prime}_i)]^2 \\
  & \text{subject to}
  & & \arctan \norm{\textbf{R}} \;\in \,[0,\pi / 2),
  \end{aligned}
\end{equation}

where $v(\Omega_i^{\prime})$ represents the volume of the bin $\Omega_i$ after transformation; $\Omega_i^{\prime}$ has been translated by the rotation on $\textbf{X}$ and normalised such that $v(\Omega^{\prime})=1$. $\textbf{R}$ is as defined in equation (\ref{eq:mrp}).

The relationship between $\textbf{R}$ and $v(\Omega^{\prime})$ is non-trivial, since the size of the support is not constant under rotation. Consequently, the derivative appears not to be well-defined. However, numerical methods allow this to be approximated, and it is observed that routines converge to the global optimum.

\begin{figure}[htb]
  \centering
  \includegraphics[width=\columnwidth]{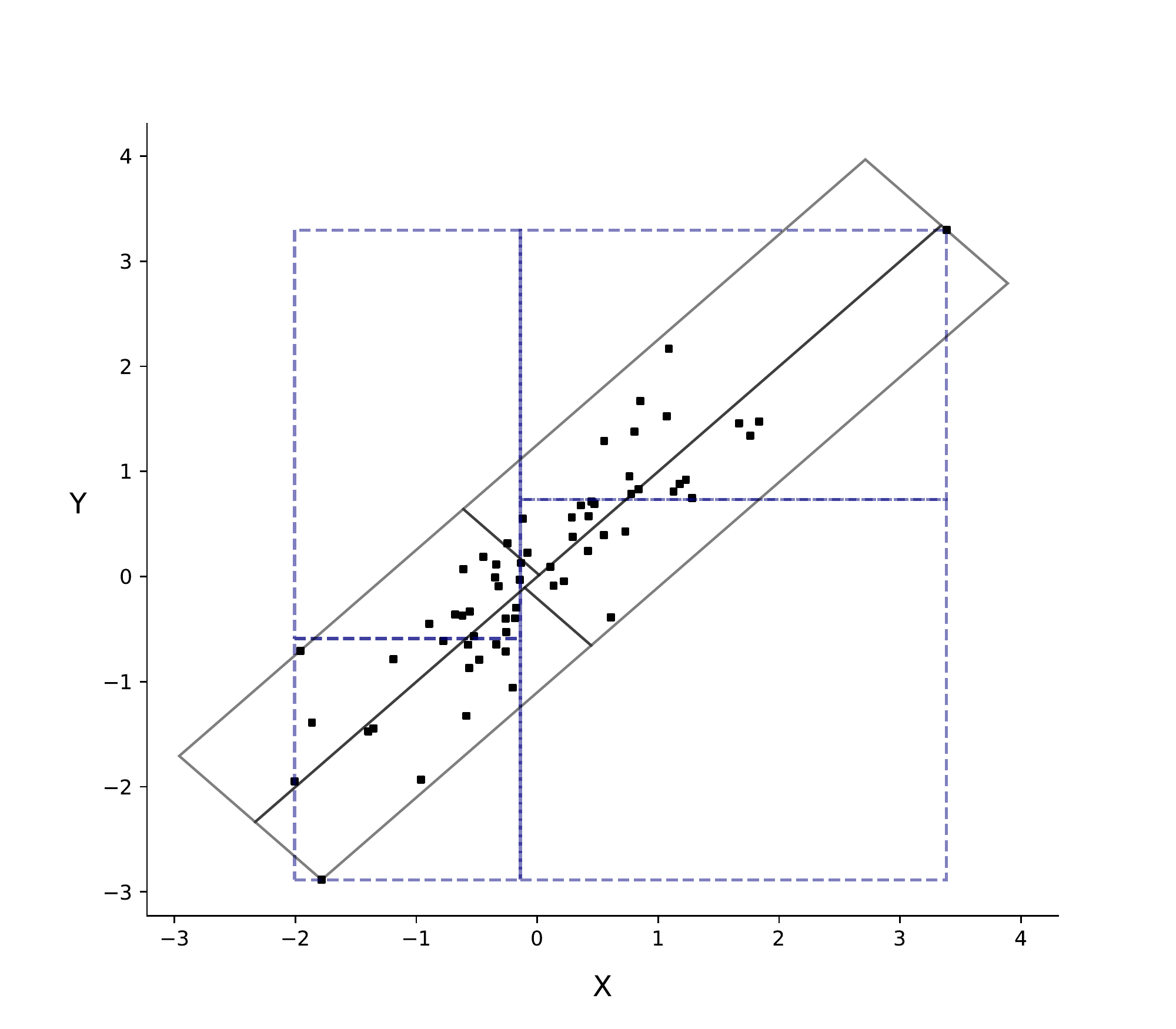}
  \caption{Synthetic data ($N=64$) randomly sampled from $Y = X + \mathcal{N} (0,0.5)$, where $X \sim \mathcal{N} (0,1)$, and two equiprobable partitions of the sample space, both calculated using recursive binary trees to depth $D=1$. The dashed-line partition follows the naive orientation of the marginals; the variance of its bin volumes is calculated to be $9.4$. The solid-line partition represents the orientation with minimum variance of bin volumes, which is calculated to be $0.5$. In this case, the optimal alignment requires a rotation of $\pi / 4$ \text{rad}. It is noted that for the purposes of visualisation the rotated partition is plotted at an angle against the original axes; in fact the axes themselves will be rotated to the new domain.} 
  \label{fig.rotation}
\end{figure}

Figure (\ref{fig.rotation}) presents two equiprobable histograms prepared for a sample of synthetic, correlated two-dimensional data. One histogram is aligned rotationally along the marginals, and the other along the orientation minimising the variance of histogram bin volumes. 

The linear relationship between the variables $X$ and $Y$ elucidates the benefit of optimal rotation. It can be seen that the marginally-aligned histogram is constrained by its alignment and effectively over-fits the data - it overestimates the variance in the bin volumes, which results in underestimation of the sample entropy. In effect, the estimator suggests more information than is truly presented by the data; the rotational alignment of the partition introduces intrinsic information from the perpendicular construction of the histogram partition itself. By optimising the rotational orientation of the partition, as demonstrated by the other histogram, this additional information can be minimised.

This linear example also prompts the corollary: that nonlinear and rotationally symmetric distributions may see only limited improvement from the rotational alignment procedure. 

Interestingly, the optimum alignment in the figure is calculated to be $\pi / 4$ rad, which observation leads to a secondary supposition: that the optimal alignment could usefully be approximated by the eigenvectors. This has computational appeal, as the optimisation routine is expensive compared to the calculation of eigenvectors (except for very large $N$), and also because the rotation operation would then naturally generalise to higher dimensions. However further research is required to provide bounds on the errors from adopting such an assumption. 

\subsection{Challenges and Limitations} \label{s.limitations} 
  The method presented in this paper improves upon existing techniques for density estimation in several important ways. By use of a recursive binary partition, I construct an equiprobable histogram and demonstrate that a plug-in entropy estimator can be calculated from the relative volumes of the leaf nodes. I demonstrate that the rotational alignment of the histogram to the marginals is not required here, and conjecture that the optimal alignment does not lie along the marginals. I express this as a rotational alignment parameter to be optimised, and propose that the alignment is optimal which minimises the variance of the bin volumes. 
  
  However, there remain some challenges which must be taken into account when applying this method in practice. First, in common with other nonparametric density estimators, the sample space is dictated by the range of the sample data which gives an oversize impact of outlier points on the location and shape of the partition. For example, Stowell \& Plumbley limit the volume of the edge bins to the Maximum Likelihood Estimate of the support, which in fact is equivalent to enclosing the partition over the sample space \cite{stowell2009fast}. This complication is particularly problematic when estimating entropy in terms of relative bin volumes, as the sample space is not well defined, and the partition will be stretched along the dimensions most affected by extrema, impacting the volumes of some bins even when the extrema lie in other bins. 
  
  The optimal rotational alignment may present a natural, albeit partial redress of this problem. In some cases it may also appropriate to perform some degree of careful outlier transformation. By way of winsorisation, the extreme values of outliers could be brought within a reasonable range such as $3\sigma$ along each marginal. This would improve the robustness of the technique to outliers \cite{ruppert2014trimming}. 

  It is also interesting to note that the cycle order in which dimensions are bisected impacts the final structure of the histogram. The result is that multiple equiprobable partitions can be produced within the same domain after the optimal rotational alignment has been found. This in fact represents an opportunity for ensemble learning, by cycling through bisections beginning from each of the initial dimensions, or selecting the order at random for each depth of the recursion, and taking an average of multiple entropy estimates.

  Finally, it must be remarked that 
  the entropy equation (\ref{eq:entropy-estimate}) holds exactly $\operatorname{iff} N/m \in \mathbb{N}$, and entropy estimates calculated when this does not hold must include some inevitable variance due to this inequality. Any impact of this is generally reduced by increasing the sample size and reducing the number of bins. This suggests an open question on selecting the depth parameter $s$. However, following equation (\ref{eq:numBins}), the number of bins scales by $2^d$ for every increment to $s$, and so the decision is in practice often not so hard. A sensible rule is for there to be always at least a few samples per bin; for common sample sizes measured, in hundreds or thousands, $s$ should be at most $3$ or $4$ for bivariate data.

\section{Results} \label{s.validation}

  The performance of an entropy estimator was measured by comparing estimates over many realisations of multivariate distributions with known analytical entropies. Samples were drawn from bivariate Gaussian distributions with randomised scale parameters, designed to produce a diverse study of distributions with varying entropies and degrees of correlation. 

  For multivariate Gaussian distributions the entropy in bits can be calculated analytically from the probability distribution function, and is known to be:
  
  \begin{equation}
    \label{eq:gaussian-entropy}
    H(\textbf{X}) = \frac{1}{2} \log_2 (2 \pi e |\Sigma|) ,
  \end{equation}

  where $|\Sigma|$ represents the determinant of the covariance matrix between the marginal dimensions. The covariance matrix describes the extent to which the marginal distributions are correlated, and hence characterises the shape of the distribution; the eigenvectors point along the direction of maximum variance. Specifically, this parameterises the degree of rotational symmetry in the distribution.

  The impact of rotational alignment of the partition on the accuracy of the entropy estimate can be observed only when the sample data displays rotational asymmetry. For bivariate Gaussian distributions, this requires that samples were drawn from distributions with large covariance. To ensure this, each realisation was sampled from a bivariate Gaussian distribution with scale parameter described by a randomised covariance matrix defined by:

  \begin{equation}
    \label{eq:covariance-matrix}
    \Sigma = \begin{pmatrix}
      \mathcal{U}(-20,20) & \mathcal{U}(0,30) \\
      \mathcal{N}(-2,2) & \mathcal{N}(-1,1)
  \end{pmatrix}
  \end{equation}

  such that samples with varying degrees of rotational asymmetry could be estimated in each study.
 
  For each sample, the entropy was estimated using two different equiprobable partitions prepared using the k-d tree approach described in Section \ref{s.method}. The first partition was aligned to the marginals, whilst the second was aligned with the optimal orientation. This optimal orientation was calculated by optimising over $\textbf{R}$, following equation (\ref{eq:optimisation}), using a sequential least squares programming routine \cite{kraft1988software}. Accordingly, the effect of the rotation was able to be observed independently; following intuition, the rotated partition was expected to perform on average at least as well as the the unrotated partition. Finally, two other techniques - a naive partition of equal-width bins and a marginal-equiquantised partition - were used to estimate the entropy for comparison. Together these estimates were then compared against the theoretical entropies, and the results for each are presented in Table I. The accuracy of each technique was compared using a loss function of the absolute percentage error between the estimated entropies by each technique, and the theoretical entropies from equation (\ref{eq:gaussian-entropy}). This was aggregated across all realisations by taking the mean squared error. By adopting this loss function, relative overestimation and underestimation were treated equally, and there was no bias in favour of samples with smaller absolute entropy.

  It was observed that the naive histogram was second only to the rotated partition in mean performance, suggesting that rotational alignment is the critical component driving the performance of the optimal partition. Following this, one-sided 99\% confidence intervals were generated by bootstrapping \cite{efron1994introduction} the difference in MSE between the naive and rotated equiprobable techniques. As shown in Table I, the confidence intervals around these estimates are strictly greater than zero for each study. Hence it can be concluded that the optimal partition outperforms the naive partition in estimating multivariate entropy. 

  \begin{table*}[!htb]
    \label{t.gaussian-entropies2d}
    \caption{Mean entropy estimates by candidate partitioning techniques. Each study is defined by number of samples $N$ and associated bin counts $B$, and is calculated over 1000 observations of synthetic bivariate Gaussian distributions with varying degrees of covariance. The equiprobable partition subject to optimal rotation has the least MSE, providing evidence for the superiority of the technique. One-sided bootstrap confidence intervals, calculated for the difference in MSE between naive and optimal equiprobable partitions, are positive for all studies, demonstrating that the observed mean outperformance is significant across sample sizes.}

    \begin{tabular*}{\textwidth}{l@{\extracolsep{\fill}}lrrrrrr}
      N &  $B$   &  $\operatorname{MSE_{Naive}}$ &  $\operatorname{MSE_{Marg. Eq.}}$ & $\operatorname{MSE_{Equiprob.}}$ & $\operatorname{MSE_{Rot.Equiprob.}}$ & \makecell{ 99\% Bootstrap CI \\ ($MSE_{\operatorname{Naive}} - MSE_{\operatorname{Rot.Equiprob.}}$)} \\ 
      \hline
      32   &               4 &              0.004 &                0.015 &                                 0.011 &                     0.000 &     0.001 \\
      50   &               4 &              0.022 &                0.033 &                                 0.040 &                     0.000 &     0.004 \\
      64   &               4 &              0.020 &                0.022 &                                 0.063 &                     0.000 &     0.004 \\
      100  &              16 &              0.018 &                0.008 &                                 0.011 &                     0.002 &     0.001 \\
      300  &              16 &              0.010 &                0.011 &                                 0.025 &                     0.000 &     0.001 \\
      512  &              16 &              0.024 &                0.020 &                                 0.034 &                     0.001 &     0.001 \\
      1024 &              16 &              0.005 &                0.007 &                                 0.027 &                     0.001 &     0.001 \\

    \end{tabular*}
    \end{table*}

\section{Conclusion} \label{s.conclusions}

This paper presented a reformulation of the histogram entropy estimator in terms of relative bin volumes, alongside a novel k-d tree implementation of an equiprobable histogram partition. Demonstrating thereby that the alignment of such an equiprobable partition with the marginals is not required, it was conjectured that improved equiprobable partitions could be found by rotationally aligning the partition with the sample distribution. Such rotation operations would have the effect of reducing the extraneous information introduced to the estimate by the construction of the partition itself.

To test the conjecture, synthetic data was sampled from realisations of bivariate Gaussian distributions, generated using randomised covariance matrices specified so as to ensure rotational asymmetry. For each sample, a sequential least squares programming routine was used to find the partition with the minimum variance of bin volumes. Entropy estimates were then calculated for each sample using the optimally-aligned and the marginally-aligned equiprobable partitions, as well as two other histograms for comparison. These entropy estimates were validated against known theoretical entropies, and it was observed that the optimal partitions demonstrated significantly reduced mean square error versus the competing techniques. 

The results present strong evidence to accept the original conjecture; equiprobable histogram entropy estimates for small and highly correlated samples can be improved by aligning the partition to the optimal rotational orientation. Since, as demonstrated in Section \ref{s.background}, the orientation of the histogram bins is not relevant to the entropy, applying the rotation effectively reduces a bias towards samples which happen to be oriented in the privileged marginal reference frame. It is therefore concluded that rotational alignment should be considered a parameter of the histogram estimator which can be learned from the data; the optimal partition will perform on average at least as well as an equiprobable partition which is oriented along the marginals.

It is noted that there are some limitations to the technique, and the optimisation routine can be computationally expensive. The worth of calculating the optimal partition may be limited to problems in which precision is more important than speed in calculating the entropy estimate. This compromise could be improved by a solution for the optimal rotational alignment, which would allow the parameter to be selected without performing the numerical optimisation. 
A question left for future research is whether the optimal rotational alignment can be solved analytically; either by deriving the objective and its gradients for the recursive binary tree, or demonstrating that the optimal orientation may be bounded within some region around the sample eigenvectors. This could offer considerable improvements in the computational efficiency of the technique, particularly for estimating the entropy of high-dimensional data, as eigenvectors can be calculated efficiently for multiple dimensions whilst the optimisation complexity increases exponentially with dimensionality.

\vspace{5px} 

\section{Acknowledgments}
I would like to thank Prof. Tomaso Aste for several productive discussions on defining optimality, and for providing valuable feedback and recommendations during reviews of early drafts of this manuscript. I remain extremely grateful for his mentorship during my postgraduate study and for motivating me to pursue research in the fields of information theory and computational statistics.

\appendix

\vspace{10px} 


\bibliography{bibliography}

\end{document}